\documentclass{llncs}
\usepackage{listings}
\usepackage{amssymb}
\usepackage{amsmath}
\usepackage{color}
\usepackage{graphicx}
\begin{document}

\lstset{language=C,basicstyle=\small}
\lstset{numbers=left, numberstyle=\tiny, stepnumber=1, numbersep=5pt}
\lstset{firstnumber=1}
\lstset{frame=single}
\lstset{showstringspaces=false}
\lstset{showspaces=false}
\lstset{showtabs=false}
\lstset{tabsize=2}
\lstset{
  language={C},
  morekeywords={assert,uchar}
}

\newcommand{\newbf}[1]{\textcolor{red}{#1}}
\newcommand{\newrb}[1]{\textcolor{blue}{#1}}
\newcommand{\blurb}[1]{{\texttt{... #1 ...}}}

\title{Verifying Embedded C Software with Timing Constraints using an Untimed Model Checker}
\author{
  Raimundo Barreto \inst{1},
  Lucas Cordeiro   \inst{2}, \and 
  Bernd Fischer    \inst{3}
}

\institute{ 
  \email{rbarreto@dcc.ufam.edu.br}\\
  Dept.\ of Computer Science, Federal University of Amazonas, Manaus AM, Brazil
\and
  \email{lucascordeiro@ufam.edu.br}\\
  Dept.\ of Elect.\&Telecom., Federal University of Amazonas, 
  Manaus AM, Brazil
\and
  \email{b.fischer@ecs.soton.ac.uk}\\
  Electronics and Computer Science, University of Southampton,
  Southampton, UK
}
\maketitle

\begin{abstract}
Embedded systems are everywhere, from home appliances to critical systems such
as medical devices. They usually have associated timing constraints that need to be
verified for the implementation.  Here, we use an untimed
bounded model checker to verify timing properties of embedded C programs.
We propose an approach to specify discrete time timing constraints using code annotations. The
annotated code is then automatically translated to code that manipulates
auxiliary timer variables and is thus suitable as input to conventional, untimed software
model checker such as ESBMC.  Thus, we can check timing
constraints in the same way and at the same time as untimed system
requirements, and even allow for interaction between them.  We applied the
proposed method in a case study, and verified timing constraints of a pulse
oximeter, a noninvasive medical device that measures the oxygen
saturation of arterial blood.
\end{abstract}

\section{Introduction}

Model checking is an automatic technique for verifying finite state concurrent systems~\cite{clarke00}. 
The main problem in model checking is the well-known state space explosion;
adding real-time aspects to model checking only makes this problem worse.
Usually, real-time systems are modeled by timed automata, timed Petri net, or a kind of labeled state graphs,
and verified with specialized timed model checking tools, 
such as TINA~\cite{Tina06}, HyTech~\cite{HyTech97}, Kronos~\cite{kronos97}, or UPPAAL~\cite{upaal97}.
For example, UPPAAL uses timed automata as input and
a fragment of the TCTL temporal logic~\cite{Wang06}
to prove a safety property 
in an explicit-state model checking style. 
Here, we propose a different approach.
In our method, the safety property is specified in an explicit-time style~\cite{Lamport05}, using discrete-time timing annotations 
in ANSI-C programs. 
We assume that timing annotations are given externally, either be 
a WCET analysis of the code, or by a domain expert.
We then translate such annotated C code automatically
to code that manipulates auxiliary timer variables. This code is suitable as 
input for a conventional (i.e., untimed) software model checker; since we are 
working with a discrete-time model, timing assertions can simply be interpreted
as integer constraints.

In our implementation, we use ESBMC~\cite{cordeiro11}, a bounded symbolic
model checker for ANSI-C which is based on \emph{satisfiability modulo
theories} (SMT) techniques, while specialized timed model checkers typically
adopt an explicit-state style (e.g., UPPAAL). 
Symbolic model checkers 
can typically explore more states than explicit-state model checkers,
despite some state-space reduction techniques.
Moreover, symbolic model checking can easily be combined with powerful
symbolic reasoning methods such as decision procedures and SMT solving. 
This reduces not only the state space but also allows us to handle timing
constraints symbolically yet precisely. 
Note that the timing annotations need to be treated separately from the other
assertions during loop unrolling (which is a crucial step in \emph{bounded}
model checking) in order to get correct results. We avoid this
problem by annotating only function definitions.

Many safety-critical software systems are written in low-level languages such as
ANSI-C. However, to the best of our knowledge, there is at present no
tool that translates C code with timing constraints to either timed automata or
timed Petri nets.
The main aim of this paper is thus to propose a method to check timing properties directly in the actual C code 
using a (conventional) software model checker;
however, we can check timing properties as well as safety and liveness properties (see~\cite{cordeiro11}).
The proposed solution should not be considered as an alternative to other methods, but rather as complementary. 
There are at least two scenarios in which it can be used: 
(1) for legacy code that does not have a model, or where there 
are no automated tools to extract a faithful model from the code; and 
(2) when there is no guarantee that the final code is in strict accordance with the model. 

We focus on time-critical embedded systems software which, due to predictability issues, 
require guarantees that in all execution paths
the timing constraints are met.
We illustrate our approach through an industrial case study
involving a medical device called pulse oximeter.  Our experiments show
that our technique can be used efficiently for verifying embedded real-time systems
using an existing untimed model checker.

The main contribution of this work is to check timing properties in the same way as for untimed systems. Specifically: 
we use code annotation to express timing properties; 
we describe our 
translation from the annotated code to a code suitable for model checking; and 
we report experiments
on a medical device.

The paper is organized as follows. 
The next section shows the background to understand the proposed method.
Section 3 describes the proposed method.
Section 4 analyzes the pulse oximeter case study.
Section 5 reviews related work.
Finally, Section 6 summarizes the paper and explains future work.

\section {Backgroud}

\subsection{Model Checking with ESBMC}

ESBMC is a context-bounded model checker for embedded ANSI-C software based on SMT solvers, which allows the verification of single- and multi-threaded 
software with shared variables and locks~\cite{Cordeiro09,cordeiro11}, although we have focused in single-threaded software here.
ESBMC supports full ANSI-C, and can verify programs that make use of bit-level, arrays, pointers, structs, unions, 
memory allocation and fixed-point arithmetic. It can efficiently reason about arithmetic under- and overflows, pointer safety, 
memory leaks, array bounds violations, atomicity and order violations, local and global deadlocks, data races, 
and user-specified assertions.

In ESBMC, the program to be analyzed is modelled as a state transition system $M = (S, R, s_{0})$, which is extracted from the control-flow graph (CFG). 
$S$ represents the set of states, $R \subseteq S \times S$ represents the set of transitions 
(i.e., pairs of states specifying how the system can move from state to state) 
and $s_{0} \subseteq S$ represents the set of initial states. 
A state $s \in S$ consists of the value  of the program counter \emph{pc} and the values of all program variables. 
An initial state $s_{0}$ assigns the initial program location of the CFG to \emph{pc}. 
We identify each transition $\gamma=(s_i,s_{i+1}) \in R$ between two states $s_{i}$ and $s_{i+1}$ 
with a logical formula $\gamma(s_i,s_{i+1})$ that captures the constraints on the corresponding values 
of the program counter and the program variables.

Given the transition system \textit{M}, a safety property $\phi$, a context bound $C$ and a bound $k$, 
ESBMC builds a reachability tree (RT) that represents the program unfolding for $C$, $k$ and $\phi$. 
We then derive a VC $\psi^{\pi}_{k}$ for each given interleaving (or computation path) $\pi = \{\nu_1,\ldots, \nu_k\}$ such that $\psi^{\pi}_{k}$ 
is satisfiable if and only if $\phi$ has a counterexample of depth \textit{k} that is exhibited by $\pi$.\ $\psi^{\pi}_{k}$ is given by 
the following logical formula:

\begin{equation} \label{bounded-model-checking} \psi^{\pi}_{k} =
I(s_{0})
\wedge
  \bigvee^{k}_{i=0} \bigwedge^{i-1}_{j=0} \gamma(s_{j},s_{j+1})
\wedge
  \neg \phi(s_i)
\end{equation}

Here, $I$ characterizes the set of initial states of $M$ and $\gamma(s_{j},s_{j+1})$ is the transition relation of $M$ 
between time steps $j$ and $j+1$. Hence, $I(s_0)\wedge\bigwedge^{i-1}_{j=0} \gamma(s_{j},s_{j+1})$ represents 
executions of $M$ of length $i$ and $\psi^{\pi}_{k}$ can be satisfied if and only if for some $i \leq k$ there exists 
a reachable state along $\pi$ at time step $i$ in which $\phi$ is violated. $\psi^{\pi}_{k}$ is a quantifier-free 
formula in a decidable subset of first-order logic, which is checked for satisfiability by an SMT solver. 
If $\psi^{\pi}_{k}$ is satisfiable, then $\phi$ is violated along $\pi$ and the SMT solver provides a satisfying assignment, 
from which we can extract the values of the program variables to construct a counter-example. 
A counter-example for a property $\phi$ is a sequence of states $s_{0}, s_{1},\ldots, s_{k}$ 
with $s_{0} \in S_{0}$, $s_{k} \in S$, and $\gamma\left(s_{i}, s_{i+1}\right)$ for $0 \leq i < k$. 
If $\psi^{\pi}_{k}$ is unsatisfiable, we can conclude that no error state is reachable in $k$ steps or less along $\pi$. 
Finally, we can define $\psi_{k} = \bigwedge_{\pi}\psi^{\pi}_{k}$ and use this to check all paths.

\subsection{Model Checking Real-Time Systems}

Model checking is a verification technique that applies to systems that can be modeled by a mathematical 
formalism (finite automata and Petri nets are examples). 
In practice, the size of the systems is really the main obstacle to overcome.
Therefore, model checker users usually simplify the model under analysis.
Model checking consists in three steps.
(1) Mathematical representation (modeling). 
Usually such models represents states and transitions, and may be composed by and synchonized by several components. 
(2) Representation of a property by a logical formula. 
Once the model is built, we formally state the properties to be checked, usually in a temporal logic.
(3) Model checking algorithm. 
Given a model $\mathbf{A}$ and a property $\phi$, a model checking algorithm answers the question: 
``does the model $\mathbf{A}$ satisfy the property $\phi$?"

There are several tools that model check real-time systems.

TSMV~\cite{Markey03} is a symbolic model checker that verifies TCTL formulas on Timed Kripke Structures (TKS), 
i.e. finite state graphs where transitions carry a duration. 
The main feature of TKS's is that the durations of transitions are atomic, that is, 
when moving from state $s$ to state $s{'}$ in a step that lasts 10 time units, 
there is no intermediary configuration between $s$ at time $t$ and $s{'}$ at time $t+10$. 
The key motivation for this semantics is that it leads to simple and efficient model checking algorithms.

UPPAAL~\cite{upaal97} allows one to analyze networks of timed automata with binary synchronization. 
It contains three main parts: (i) a graphical editor where timed process are described; 
(ii) a simulator where it is possible to choose a sequence of transitions, 
and to see the behavior of the system; and
(iii) a verifier of reachability properties. 
The main drawbacks of UPPAAL are: (1) the binary synchronization 
is a bit restrictive and requires one to use ad hoc mechanisms 
to describe other kinds of synchronizations (e.g., broadcast); 
(2) the specification language considers only reachability properties and not a full temporal logic.
This entails that it is necessary to include a observer automata to express complex properties. 

KRONOS~\cite{kronos97} is a model checker that can decide whether some property, 
expressed by a TCTL formula, holds for a timed automaton 
(also called timed graph), given in textual form. 
It allows one to verify liveness properties, 
and is not restricted to reachability properties. 
Even though KRONOS contains no graphical nor simulation modes,
it is a true timed model checker. 
However, under its current form, it is mostly intended for advanced users with a good knowledge of formal methods.

HYTECH~\cite{HyTech97} receives a set of linear hybrid automata, and synchronizes them by some common transitions. 
From the automata in a textual form, HYTECH can compute subsets of the global state space. 
HYTECH also handles parametric analysis, that is, when a system (or a property) contains parameters, 
the analysis can provide the parameter values for which the property holds. 
The drawbacks of HYTECH is that it includes no simulation mode; model checking does not apply to a temporal logic: 
the user has to build himself the subset of states to be computed by combinations of basic constraints. 

TINA~\cite{Tina06} is a toolbox for the edition and analysis of Petri Nets and Time Petri Nets. 
The Tina toolbox includes several tools such as: (i) an editor for graphically or textually described 
Time Petri nets; (ii) construction of reachability graphs where it may build coverability graphs, 
persistent sets, state class graphs, and (iii) a state/event LTL model checker that checks reachability properties.
However, it is difficult to check real-time quantitative properties.

\section {Proposed Method}

\begin{figure*}[!t]
	\centering
	\includegraphics[scale=0.32]{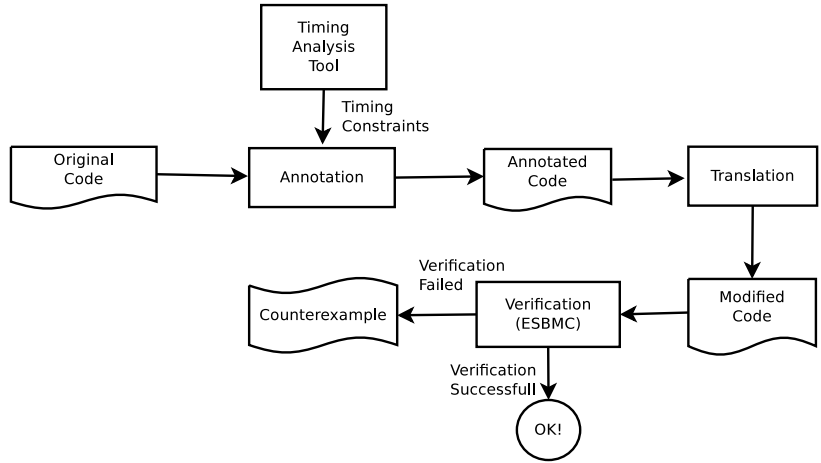}
	\caption{Overview of the Proposed Method}
	\label{fig:overview}
\end{figure*} 

This section describes the method proposed to verify timing properties on single-threaded 
C code using a bounded model model-checker.
Figure~\ref{fig:overview} gives an overview of the approach. 
It is divided into four phases. The first step is to add timing constraints to the source code. 
Such annotations come from either a discrete timing model, a timing analyzer tool, or a domain expert.
As usual, the annotations are just comments that are processed by a specific tool. 
The second step is the automatic translation from the annotated source code to new code that can be verified by the untimed model-checker. 
Basically, this translation consists in (i) adding declarations of the timer variables; (ii) gathering the annotated timing constraints 
information and including assignment statments for the new added variables; (iii) adding (user-defined) assert statements at 
specific points of the program code. 
The third step is to check the translated code with the ESBMC model checker. 
Finally, the last phase, evaluates ESBMC's results.
As shown, the way to check timing properties is by using assertion.

\subsection{Timed Programming Model}
\label{section:overview}

The proposed method aims to pragmatically assist developers in the specification and analysis of timing constraints 
in C code.
What we propose is 
(i) to associate with each function $f_i$ a worst-case duration $d_i \geq 0$; 
(ii) to define explicit timer variables (or clocks) ($\mathcal{T}$), 
for expressing timing constraints; 
(iii) to introduce timing assertions on timer variables to check timing properties; and
(iv) to introduce  timer variable {\it reset} to restart the timer counting.
Therefore, when the program is executed, the timer variables are incremented by the respective 
duration $d_i$ of the called function $f_i$, and
assertions are used to ensure that computations are within timing constraints.

Formally,
let consider that the semantics of a sequential program $\mathcal{P}$ is represented by the {\it 5}-tuple 
$\langle \mathcal{S}, s_0, \mathcal{V}, \mathcal{F}, \rightarrow \rangle$, 
where:
\begin{itemize}
\item $S$ is a finite set of states of $\mathcal{P}$;
\item $s_0$ is the initial state;
\item $\mathcal{V} = \langle v_1, v_2, \cdots, v_z\rangle$ is a finite list of data variables (local and global);
\item $\mathcal{F} = \{f_1, f_2, \cdots, f_w\}$ is a finite set of functions that may change variables
in $\mathcal{V}$; 
\item $\rightarrow \subseteq \mathcal{S} \times \mathcal{F} \times \mathcal{S}$ is a finite set of labeled transitions,
such that a state transition in $\langle s_i, f_{\phi}, s_j\rangle$, 
is represented by 
$s_i \stackrel{f_{\phi}} \longrightarrow s_j$, $\forall s_i,s_j \in S, i \neq j, f_{\phi} \in \mathcal{F}$ and 
it is supposed that each function $f_{\phi}$ is executed to completion.
\end{itemize}

Let $\pi[n \ldots m]$ for $0 \leq n < m \in \mathbb{N}$ be an {\it execution path} denoted by a finite sequence 
$s_n \stackrel{f_{\phi_{1}}} \longrightarrow s_{n+1} \stackrel{f_{\phi_{2}}}  \longrightarrow \cdots
\stackrel{f_{\phi_{q}}} \longrightarrow s_{m}$
with $m-n$ transitions and $m - n + 1$ states. 
As example, suppose we have
$\mathcal{F} = \{f_1, f_2, f_3, f_4, f_5\}$;
we may define the following execution path
$\pi [0..5]= s_0 \stackrel{f_1} \longrightarrow s_1 \stackrel{f_3} \longrightarrow 
s_2 \stackrel{f_5} \longrightarrow s_3 \stackrel{f_2} \longrightarrow
s_4 \stackrel{f_4} \longrightarrow s_5$.

In order to introduce timing constraints into the program, we change the original 
program $\mathcal{P}$ to another program $\mathcal{P'} = 
\langle \mathcal{S}, s_0, \mathcal{V'}, \mathcal{F'}, \rightarrow \rangle$ 
where:
\begin{itemize}
\item $\mathcal{V^{'}} = $ {\tt cat}$(\mathcal{V}, \mathcal{T})$, where {\tt cat} means list concatenation
 in this case used to concatenate lists of variables;
\item $\mathcal{F^{'}} = \mathcal{F}  \cup \mathcal{A}  \cup \mathcal{R}$;
\item $\mathcal{T} = \langle t_1, t_2, \cdots, t_p \rangle$ is a finite list of timer variables;
\item $\mathcal{A} = \{ a_1(t_{k_1}), a_2(t_{k_2}), \cdots, a_n(t_{k_x}) \; |  \;
a_i$ is a special function that asserts on timer variables, $1 \leq i \leq n,$ and
 $t_{k_d} \in \mathcal{T}, 1 \leq d \leq x, 1 \leq k_d \leq p, p = | \mathcal{T} | \}$;  
\item $\mathcal{R} = \{ r_1(t_{1}), r_2(t_{2}), \cdots, r_p(t_{p}) \; |  \;
r_w$ is a special function that resets a timer variable, $1 \leq w \leq p$, $p = | \mathcal{T} |$, and $t_w \in \mathcal{T}\}$.
\end{itemize}

We define $D: \mathcal{F}^{'} \mapsto \mathbb{N}$ 
as the worst-case duration of a function, such that
$D(f_i),\;\forall f_i \in \mathcal{F}^{'} = 
\begin{cases}
d_i \in \mathbb{N}, & \mbox{if} (f_i \in \mathcal{F}) \\
0, & \mbox{if} (f_i \in \mathcal{A}) \\
0, &  \mbox{if} (f_i \in \mathcal{R}) \\
\end{cases}
$

Therefore, we may express the duration $D(\pi [n ..m]) \; = \; \sum_{i=1}^{m-n}D(f_{\phi_{i}})$ 
of such a finite sequence $\pi [n ..m]$
representing the time elapsed from $s_n$ to $s_m$.
As example, suppose we have
$\mathcal{F} = \{f_1, f_2, f_3, f_4, f_5\}$;
$\mathcal{T} = \{t_1, t_2\}$;
$\mathcal{A} = \{a_1(t_1), a_2(t_1),$ $a_3(t_2)\}$;
$\mathcal{R} = \{r_1(t_1), r_2(t_2)\}$;
and  the execution path
$\pi [0..11]= s_0 \stackrel{r_1(t_1)}   \longrightarrow s_1 \stackrel{r_2(t_2)}   \longrightarrow 
             s_2 \stackrel{f_1}      \longrightarrow s_3 \stackrel{f_3}       \longrightarrow
             s_4 \stackrel{a_1(t_1)} \longrightarrow s_5 \stackrel{r_1(t_1)}    \longrightarrow 
             s_6 \stackrel{f_5}      \longrightarrow s_7 \stackrel{f_2}       \longrightarrow
             s_8 \stackrel{a_2(t_1)} \longrightarrow s_9 \stackrel{f_4}       \longrightarrow 
             s_{10} \stackrel{a_3(t_2)}  \longrightarrow s_{11}
$, where
$f_{\phi_{1}}=r_1(t_1)$;
$f_{\phi_{2}}=r_2(t_21)$;
$f_{\phi_{3}}=f_1$;
$f_{\phi_{4}}=f_3$;
$f_{\phi_{5}}=a_1(t_1)$;
$f_{\phi_{6}}=r_1(t_1)$;
$f_{\phi_{7}}=f_5$;
$f_{\phi_{8}}=f_2$;
$f_{\phi_{9}}=a_2(t_1)$;
$f_{\phi_{10}}=f_4$;
$f_{\phi_{11}}=a_3(t_1)$.
We can conclude that
$D(\pi [0 ..11]) = \sum_{i=1}^{11} D(f_{\phi_{i}}) = D(f_1) + D(f_3) + D(f_5) + D(f_2) + D(f_4)$.
As we can see, in the execution path $\pi[0 ..11]$ we have three timing verifications: $a_1(t_1)$, $a_2(t_1)$, and $a_3(t_2)$; and 
three timer resets: $r_1(t_1), r_2(t_2),$ and $r_1(t_1)$.

\subsection{Annotation of Timing Constraints}
\label{section:annotation}

The inclusion of timing constraints in the source code is particularly interesting since it can automatically be checked as the 
program are being developed. 
To annotate the timing constraints in the code we use a special kind of C comment in such a way that this annotation does not change 
the code itself. 
In this way, the same annotated code can be compiled by any C compiler without breaking the compilation. 
The proposal is to have four kinds of annotations:
\begin{itemize}
 \item {\tt //@ DEFINE-TIMER <timer-name>}. Defines a new timer variable 
   \emph{timer-name} which is automatically declared as an unsigned int variable.
    Using this annotation we can add the set $\mathcal{T}$ to the code.
 \item {\tt //@ RESET-TIMER <timer-name>}. Resets the timer variable to zero. 
    Using this annotation we can add the set $\mathcal{R}$ to the code.
 \item {\tt //@ ASSERT-TIMER (<logic-expr>)}. Checks a user defined assert. 
  This annotation specifically is useful to check timing properties, 
where the assertion language consists in arithmetic operations with timer variables.
    Using this annotation we can add the set $\mathcal{A}$ to the code.
 \item {\tt //@ WCET-FUNCTION [<int-expr>]}. Defines the WCET of the next defined function. 
  Thus, we rely on a timing analyzer tool to predict worst-case timing bounds, for instance~\cite{Bygde09}.
    Using this annotation we can add the function $D$ to the code.
\end{itemize}
Figure 2(a) shows an example of code annotation from the example shown 
on Section~\ref{section:overview}.
Even though all timer variables are incremented together, the fact that we have defined more than one timer
implies that we may verify several timing constraints. 
In the example of Figure 2, the TIMER1 is checking local timing constraints.
Firstly, this timer verifies timing constraint related to functions $f_1()$ and $f_2()$. 
Later, this same timer is then used to verify timing constraint over the functions $f_3()$ and $f_4()$. 
On the other hand, the timer variable TIMER2 is used to verify the complete behavior of the sysem, i.e.,
the function calls from $f_1()$ up to $f_5()$.

In this paper we are just showing a coarse-grained timing constraint resolution in the level of functions. 
Therefore, we show only how to specifiy timing constraints in the source code on functions. 
However, as ongoing work, we are extending the proposed annotation method to consider fine-grained 
(in the level of instructions) timing constraints.

\begin{figure}[!htb]
\center
\begin{minipage}[b]{.45\textwidth}
\label{code:annotation-and-translation1}
{\scriptsize
\begin{verbatim}
//@ DEFINE-TIMER TIMER1;

//@ DEFINE-TIMER TIMER2;

...
//@ WCET-function [d1]
void f1(void)...
//@ WCET-function [d2]
void f2(void)..
//@ WCET-function [d3]
void f3(void)...
//@ WCET-function [d4]
void f4(void)...
//@ WCET-function [d5]
void f5(void)...
...
int main(int argc, char *argv[])
...
//@ RESET-TIMER TIMER1=0;

//@ RESET-TIMER TIMER2=0;

f1(); f2();
//@ ASSERT-TIMER (TIMER1 <= alpha);

//@ RESET-TIMER TIMER1=0;

f3(); f4();
//@ ASSERT-TIMER (TIMER1 <= beta);

f5();
//@ ASSERT-TIMER (TIMER2 <= gamma);

...
               (a)
\end{verbatim}
}
\end{minipage}
\hfill
\begin{minipage}[b]{.53\textwidth}
{\scriptsize
\begin{verbatim}
// DEFINE-TIMER TIMER1;
unsigned int TIMER1;
// DEFINE-TIMER TIMER2;
unsigned int TIMER2;
...
// WCET-function [d1]
void f1(void) {TIMER1 += d1; TIMER2 += d1; ... } 
// WCET-function [d2]
void f2(void) {TIMER1 += d2; TIMER2 += d2; ... }
// WCET-function [d3]
void f3(void) {TIMER1 += d3; TIMER2 += d3; ... }
// WCET-function [d4]
void f4(void) {TIMER1 += d4; TIMER2 += d4; ... }
// WCET-function [d5]
void f5(void) {TIMER1 += d5; TIMER2 += d5; ... }
...
int main(int argc, char *argv[])
...
// RESET-TIMER TIMER1=0;
TIMER1 = 0;
// RESET-TIMER TIMER2=0;
TIMER2 = 0;
f1(); f2();
// ASSERT-TIMER (TIMER1 <= alpha);
assert (TIMER1 <= alpha);
// RESET-TIMER TIMER1=0;
TIMER1 = 0;
f3(); f4();
// ASSERT-TIMER (TIMER1 <= beta);
assert (TIMER1 <= beta);
f5();
// ASSERT-TIMER (TIMER2 <= gamma);
assert (TIMER2 <= gamma);
...
               (b)
\end{verbatim}
}
\end{minipage}
\label{code:annotation-and-translation}
\caption{(a) Example of Annotated C Code; and (b) Translation Result}
\end{figure}

\subsection{Translation and Verification}

The translation consists in looking for comments that start by {\tt //@} and treat them appropriately. 
The translation of the code shown on Figure 2(a)  
can be seen in Figure 2(b).
This translation is carried out automatically by a specific tool\footnote{This tool is available at http://esbmc.org}. 
It is important to emphasize that the user has first to run the model checker to find conventional errors 
(e.g., buffer overflow, arithmetic overflow, memory leaks, etc), 
and then run the model check to find for timing violations in the {\it modified} code.

After translation, this new code is able to be run on ESBMC, which 
check properties using user-specified assert statement.
In the proposed method the assert will be the way to check timing properties.
In the code of Figure 2(b) we may see three timing verification.

\subsection{Verifying the Bridge Crossing Problem}
\label{section:toy_example}

The bridge-crossing problem is a mathematical puzzle with real-time aspects~\cite{Rote02}. 
Four persons, $P_1$ to $P_4$, have to cross a narrow bridge.
It is dark, so they can cross only if they carry a light. 
Only one light is available and at most two persons can cross at the same time. 
Therefore any solution requires that, after two persons cross the bridge, 
one of them returns, bringing back the light for any remaining person(s). 
The four persons have different maximal speeds: 
$P_i$ crosses in $t_i$ time units (t.u.), 
and we assume that $t_1\leq t_2\leq t_3\leq t_4$.
When a pair crosses the bridge, they move at the speed of the slowest person in the pair. 
Consider that $t_1=5$; $t_2=10$; $t_3=20$; and $t_4=25$, 
the question is: how much time is required before the whole group is on the other side?
Rote~\cite{Rote02} pointed out that the most obvious solution is to let the 
fastest person (P1) accompany each other
person over the bridge and return alone with the lamp. 
In this case, the total duration of this solution is
$t_2+t_1+t_3+t_1+t_4=2t_1+t_2+t_3+t_4=65$ t.u.
However, the obvious solution is not optimal. 
The correct solution in this case is to let P3 and P4 cross in the middle.
Hence, the new total duration is
$t_2+t_1+t_4+t_2+t_2=t_1+3t_2+t_4=60$ t.u.

We implemented this problem\footnote{The code, counterexample and explanation on the results may be downloaded at 
http://esbmc.org} and submitted it the ESBMC model checker. 
We first verified that 60 is indeed the optimal solution, i.e., that
the elapsed time cannot be less than 60. 
The timing assertion and the ESBMC's output can be seen in Figure~\ref{code:toy_result_less}.
We can see that the verification failed, which means that ESBMC find 
\emph{at least} one execution path where the asserted condition is false.
The ESBMC spent 3m25s to give the result.
The second attempt was to check if the time duration could be greater or equal to 60.
The result was SUCCESSFULL, which means that the SMT solver found that in all execution paths 
the assert condition is true.
With this two results, we may conclude that $time=60$  is indeed the optimal solution.
The ESBMC spent 16m28s to give the result.
These experiments were conducted on an Intel Pentium Dual CPU with 4 GB of RAM running Linux
OS, and ESBMC v.1.15.1 (64bits).

\begin{figure}[!htb]
{\scriptsize
{\tt
\begin{minipage}{\textwidth}
\begin{verbatim}
//@ ASSERT-TIMER (__timing__ < 60);
assert (__timing__ < 60);
...
size of program expression: 37084 assignments
Generated 1 VCC(s), 1 remaining after simplification
Encoding remaining VCC(s) using bit-vector arithmetic
Solving with SMT Solver Z3 v2.16
Runtime decision procedure: 177.843s
Building error trace
Counterexample:
...
Violated property:
  file __bridge__LT60.c line 151 function main
  assertion
  __timing__ < 60

VERIFICATION FAILED
\end{verbatim}
\end{minipage}
}
}
\caption{Verification failed as the result of the application of the ESBMC}
\label{code:toy_result_less}
\end{figure}

\section{Pulse Oximeter Case Study}

This section describes the main characteristics of the pulse oximeter and shows results on the application of the
model checker ESBMC in the verification of timing constraints.
All experiments were conducted on an otherwise idle Intel Pentium Dual CPU with 4 GB of RAM running Linux
OS. We chose ESBMC v.1.15.1-64bits as untimed bounded model checker.

\subsection{Problem Specification}

The pulse oximeter is responsible for measuring the oxygen saturation ($SpO_2$) and heart rate (HR) in the blood system using a
non-invasive method. 
This device was used as case study in~\cite{Cordeiro09} to raise the coverage of tests in embedded system combining
hardware and software components. 
The implementation is relatively complex, since the final version has approximately 3500 lines of ANSI-C code 
and 80 functions. 
%
Considering that such paper does not checked timing constraints explicitly, 
and the implementation is publicly available\footnote{Availabe at: http://esbmc.org}, 
we used this problem as a case study for our proposed method.

The architecture consists in four components: sensor, data acquisition module 
(OEM-III)\footnote{For more information refer to www.nonin.com/OEMSolutions/OEM-III-Module}, microcontroller, and LCD display.
The sensor captures data on oxygen saturation (SpO2) and heart rate (HR) of the patient.
The OEM III module has an interface for communication with sensor, an ASIC (Application-Specific Integrated Circuit) component, 
and a serial communication interface (RS-232).
The ASIC component provides the values of SpO2 and HR data in the serial port.
The microcontroller receives this data, via serial port, treat them and displays on the LCD.

The packet description of the data format is shown in Table~\ref{table:packet}.
A frame consists of 5 bytes; and a packet consists of 25 frames. Three packets (375 bytes) are transmitted each second.
In this table, Byte1 is always ``01" (usually used for synchroization); 
Byte2 is the status byte (if sensor is not connected, for instance);
Byte3 shows the 8-Bit Plethysmographic Pulse Amplitude; Byte4 presents HR and SpO2 data; 
and Byte5 is the checksum.

\renewcommand{\baselinestretch}{1.2}
\setlength{\tabcolsep}{4pt}
\begin{table}[!hbt]
\begin{center}
\caption{Packet Description}
\label{table:packet}
\begin{tabular}{| r | r | r | r | r | r | }
\hline
    \# & Byte1 & Byte2 & Byte3 & Byte4 & Byte5  \\
\hline\hline
    1 & 01 & STATUS & PLETH & HR MSB & CHK \\ 
\hline
    2 & 01 & STATUS & PLETH & HR LSB & CHK \\ 
\hline
    3 & 01 & STATUS & PLETH & SpO2 & CHK \\ 
\hline
    ... & ... & ... & ... & ... & ... \\ 
\hline
    25 & 01 & STATUS & PLETH & reserved & CHK \\ 
\hline
\end{tabular}
\end{center}
\end{table}

In the context of timing constraints, the following functional requirements are considered:
\begin{enumerate}
 \item [FR1.]The system has to read all HR and SpO2 data in at most 1 second.
       In this case, we have to take into account the maximum frequency of the serial communication (9600bps),
       and the amount of bytes sent by the sensor device.
 \item [FR2.]The software must check whether the frames sent by the sensor is correct, and show in the LCD if found any problem.
       This implies in verification of the checksum, and status bytes.
       Besides that, we have to show any problem in the LCD display.
 \item [FR3.]The user should be able to see, every second, the data of heart rate and oxygen saturation in the patient's blood.
       Therefore, we have to store patient's information and to show in the LCD display.
 \item [FR4.]The system must allow users to store data on HR and SpO2 in the external memory of the microcontroller.
       We have to consider the the amount of data and the time to store in the external memory.
\end{enumerate}

\subsection{Code Annotation}

The timing constraints for this project is shown in Table~\ref{table:deadlines}. 
These constraints come from either the specification or a domain specialist.
As presented before (see Section~\ref{section:annotation}), these timing constraints are annotated into the code. 
It is worth noting that if one function calls another function, the timing constraint may be specified 
on the caller function, or on the called function, or both.

\renewcommand{\baselinestretch}{1.2}
\setlength{\tabcolsep}{2pt}
\begin{table}[!hbt]
\begin{center}
\caption{Timing Information}
\label{table:deadlines}
\begin{tabular}{l l l r }
\hline
    ID & Function & Description & WCET($\mu$s)   \\
\hline\hline
   f1 & receiveSensorData & receives data from the sensor       & 1000 \\ 
   f2 & checkStatus       & checks status      & 700 \\
   f3 & printStatusError  & displays status error     & 10000 \\
   f4 & checkSum     & calculates checksum & 2000 \\
   f5 & printCheckSumError  & displays checksum error     & 10000 \\
   f6 & storeHRMSB         & stores HR data & 200 \\ 
   f7 & storeHRLSB         & stores HR data & 200 \\ 
   f8 & storeSpO2         & stores SpO2 data & 200 \\ 
   f9 & averageHR        & calculates average of HR data & 800 \\ 
   f10 & averageSpO2      & calculates average of SpO2 data & 800 \\ 
   f11 & getHR             & returns the stored HR value         & 200 \\
   f12 & getSpO2           & returns the stored SpO2 value       & 200 \\   
   f13 & printHR           & displays HR on the LCD & 5000 \\   
   f14 & printSpO2         & displays SpO2 on the LCD & 5000 \\   
   f15 & insertLog         & inserts HR/SpO2 in RAM microcontroller & 500 \\
\hline
\end{tabular}
\end{center}
\end{table}

\subsection{Verification Results}

The pulse oximeter code is part of a real implementation.
The code adopted, and the verification results are publicly available at: {\bf http://esbmc.org}. 
In order to verify the timing constraints using ESBMC, we had to isolate hardware-dependent code. 
With this aim we used {\tt \#if}, {\tt \#else}, and {\tt \#endif} preprocessor directives.
This experiment verifies if in all execution paths the timing constraints are met 
when implementing the four functional requirements ($FR_1$, $FR_2$, $FR_3$, and $FR_4$). 
This program behavior is explained as follows:
The specification considers that we should read three packets of data per second. 
Each packet has twenty five frames.
Each frame has five bytes.
In this way we have to:
\begin{enumerate}
  \item read data bytes calling function $f_1$ ({\tt receiveSensorData});
  \item for each byte read:
  \begin{enumerate}
    \item to check status of the second byte of each frame by calling function $f_2$ ({\tt checkStatus}); 
    if there is an error, it should be called the function $f_3$ ({\tt printStatusError});
    \item to check the fifth byte of each frame by calling function $f_4$ ({\tt checkSum}); 
    if there is an error, it should be called the function $f_5$ ({\tt printCheckSumError});
    \item to read the fourth byte of first frame and to call function $f_6$ ({\tt storeHRMSB});
    \item to read the fourth byte of second frame and to call function $f_7$ ({\tt storeHRLSB});
    \item to read the fourth byte of third frame and to call function $f_8$ ({\tt storeSpO2});
  \end{enumerate}
  \item call the functions $f_9$ ({\tt average\_HR}), $f_{10}$ ({\tt average\_SpO2}), $f_{11}$ ({\tt getHR}), $f_{12}$ ({\tt getSpO2}), 
$f_{13}$ ({\tt printHR}) with HR value as argument, $f_{14}$ ({printSpO2}) with SpO2 value as argument, 
$f_{15}$ ({insertLog}) with HR value as argument, and $f_{15}$ ({insertLog}) with SpO2 value as argument.
\end{enumerate}

\linespread{0.7}
\begin{figure}[!htb]
\centering
{\footnotesize
\begin{minipage}{\textwidth}
\begin{verbatim}
...
// DEFINE-TIMER TIMER;
unsigned int TIMER;
...
//@ WCET-FUNCTION [5000]
void printHR (unsigned int line, unsigned int valueHR)
{
TIMER += 5000;
    char sHR[16];
    sprintf(sHR, "HR:%d\n", valueHR);  
    printLCD(sHR, line, 1);
}
...
int main(void) {
...
// RESET-TIMER TIMER;
TIMER=0;
...
  for (k=0; k<3; k++) {
    for (j=0; j<25; j++) {
      for (i=0; i<5; i++) {
        Byte[i] = receiveSensorData();
        if ((i==1) && (checkStatus(Byte[i])))
          printStatusError(LINE1);
        if ((i==4) && (checkSum(Byte)))
          printCheckSumError(LINE2);
        if (i==3) {
          if (j==0) storeHRMSB (Byte[i], k);
          if (j==1) storeHRLSB (Byte[i], k);
          if (j==2) storeSpO2  (Byte[i], k);
        }
      }
    }
  }

  averageHR();   
  averageSpO2();
  HR = getHR();
  SpO2 = getSpO2();
  printHR(LINE1, HR);
  printSpO2(LINE2, SpO2);
  insertLog(HR);
  insertLog(SpO2);

  // ASSERT-TIMER (TIMER < 1000000)   // one second;
  assert (TIMER < 1000000);
  ...
}
\end{verbatim} 
\end{minipage}
} 
\caption{Code for running in the ESBMC model-checker}
\label{experiment1} 
\end{figure}

\renewcommand{\baselinestretch}{1.2}
\setlength{\tabcolsep}{4pt}
\begin{table}[!hbt]
\begin{center}
\caption{Experimental Results}
\label{table:results}
\begin{tabular}{| r | r | r | r |}
\hline
    ID & \% Checksum Error & Time(s) & Result  \\
\hline\hline
    1 & 0\% & 28.9 & successful \\ 
\hline
    2 & 16.6\% & 20.3 & successful \\ 
\hline
    3 & 20\% & 20.2 & successful \\ 
\hline
    4 & 25\% & 19.9 & successful \\ 
\hline
    5 & 33.3\% & 19.9 & successful \\ 
\hline
    6 & 50\% & 21.1 & failed \\ 
\hline
    7 & 100\% & 30.2 & failed \\ 
\hline
\end{tabular}
\end{center}
\end{table}

Figure~\ref{experiment1} depicts part of the pulse oximeter code submitted to the ESBMC. 
Table~\ref{table:results} shows the experimentl results. 
We experimented seven scenarios taking into account the percentage of checksum errors.
The percentage considered was 0\%, 16.6\%, 20\%, 25\%, 33.3\%, 50\%, and 100\%.
Excepting the best scenario (0\%) and worst scenario (100\%), all timing performance was 20s in average.
As presented in Table~\ref{table:results}, if considered the worst-case scenario, 
in this case 100\% of data error, timing constraints are not met. 
However, if it is considered that it is not practical to consider an extreme situation like this one, 
we may conclude that up to 33.3\% of checksum errors, 
the system will continue to reach the timing constraints.

\section{Related Work}

Lamport~\cite{Lamport05} advocates that most real-time specifications can be verified using existing languages 
and methods. 
He proposed to represent time as an ordinary variable (now), which is incremented by an action (Tick), 
and express timing requirements 
with a special timer variable in such a way that such specifications can be verified 
with an conventional model checker. 
He call this method as model checking explicit-time specifications. 
He proposes to specify the system and timing constraints using TLA+ (Temporal Logic of Actions),
which is a high-level mathematical language. 
The problem is that the learning curve of the TLA+ may be high.

Ostroff and Hg~\cite{Ostroff96} presented a framework that allows specification, development and verification 
of discrete real-time properties of reactive systems. This framework considers a Timed Transition Model (TTM)
as underlying computational model, and Real-Time Temporal Logic (RTTL)
as the requirements specification language.
The authors provide a conversion procedure for mapping the model and specification into a finite state fair
transition systems, which may be input to a (untimed) tool, in this case STeP model-checker~\cite{Bjorner00}, 
for state exploration for checking real-time systems properties.
One problem of this method is that the converted clocked formulas count the number of tick events that occur. 
Thus, the size of the formulas grow according to the bounds that must be checked.
Since the cost of checking a linear time formula is exponential in the size of 
the formula, these procedures are only useful for small bounds.

Chun and Hung~\cite{Chun04} propose a new class of Duration Calculus (DC) called $DC^{*}_{\leq1}$, 
whose formulas correspond to expressions over the set of state occurrence for one time unit (or less),
and using conventional variables to implement relative time. 
They model the real-time system using DC and convert its components into $DC^{*}_{\leq1}$ specifications. 
Each $DC^{*}_{\leq1}$ specifications is translated to an automata model.
In this way, the whole system is modeled by the synchronisation of several automatas. 
Later, the resulting automata is translated to the SPIN untimed model checking language (in this case Promela). 
They applied their method in the Biphase Mark Protocol. 
However, they do not show how translate from automata model to Promela language. 
Additionaly, it is not clear what timing constraint was verified in this case study.

Ganty and Majumdar~\cite{Ganty09} show that checking safety properties for real-time event-driven programs
is undecidable. The undecidability proof for the safety checking problem uses an encoding 
of the execution of a 2-counter machine as a real-time event-driven program. 
The result is undecidable because such programs are not finite-state.
In this case, the task buffer as well as the call stack can grow unboundedly
large in the course of the execution.
They suggest to use higher-level languages, such as Giotto, 
which statically restricts the ability to
post tasks arbitrarily, these higher-level languages ensure that for any 
program, at any point of the execution, there is at most a bounded, statically determined, 
number of pending calls. In this case, just by finiteness of the
state space, all verification problems are decidable.

The input of the related work analyzed are: Temporal Logic of Actions (TLA) specifications, 
Timed Transition Model + Real-Time Temporal Logic (TTM/RTTL), 
Duration Calculus (DC), and Giotto programs. 
To the best of our present knowledge, there is no work that verify timing constraints using 
C code as input language.

\section{Conclusions and Future Work}

Model checking is often used for finding errors rather than for proving that they do not exist.
However, model checkers are capable of finding errors that are not likely to be found by simulation or test.
The reason for this is that unlike simulators/testers, which examine a relatively small set of
test cases, model checkers consider all possible behaviors of the system.

This paper described how to use an untimed software model checker
to verify timing constraints in C program language. 
In our proposed method we use the C language because it is one of the most commom implementation language 
of embedded systems.
As far as we are aware, there are only few approaches that deal with model-checking 
timing constraints using C code as input language.
We specified the timing behavior using an explicit-time code annotation technique  
for verifying timing properties using ordinary model checkers.
The main advantage of an explicit-time approach is the ability to use languages and tools not specially 
designed for real-time model checking.
As pointed out by Lamport~\cite{Lamport05} ``the results reported that verifying explicit-time specifications 
with an ordinary model checker is not very much worse than using a real-time model checker".

Experimental results have shown that the proposed method is promissing, 
mainly because now it is possible to verify timing constraints in the C code. 
Therefore, we are just following a movement toward application of formal verification techniques 
to the implementation level. 
In this case, we avoid constructing models explicitly and go directly to code verification. 
As presented before, this method is particularly interesting when taking into account legacy code.
However, we argue that our proposed method is not an alternative to methods currently available in the literature, 
but complementary.
We also show that using our proposed method it is possible to investigate several scenarios.

This paper just considered single-threaded code. 
It is an ongoing work to consider multi-threaded code, which is also supported by ESBMC.
Therefore, we need to consider interleavings and priorities.

This work proposed just a coarse-grained timing constraint resolution, in this case, we considered just 
timing constraints in functions. 
Thus, one future work is to extend both the code annotation method and the timing verification 
to consider fine-grained timing constraints, maybe in critical sections, for instance 
to add timing duration between two instructions representing a sequential block, 
and timing bounds for loops.

\smallskip{{\bf Acknowledgements.}~~The authors acknowledge the support granted by
 FAPESP process 08/57870-9,  CAPES process BEX-3586/10-3, 
and by CNPq processes 575696/2008-7, and 573963/2008-8.}

\linespread{1}
\bibliographystyle{plain}
\bibliography{posdoc}

\end{document}